\documentstyle[prd,aps]{revtex}
\begin{document}
\draft

\twocolumn[
\hsize\textwidth\columnwidth\hsize\csname
@twocolumnfalse\endcsname

\title{\noindent {\small USC-97/HEP-B4\hfill \hfill hep-th/9706185}
\newline
{\small  CERN-TH/97-141 \hfill}\bigskip\newline
Duality Covariant Type IIB Supersymmetry \\
and Nonperturbative Consequences}
\author{ \bf Itzhak Bars  \medskip }
\address{\it Theory Division, CERN, CH-1211 Geneva 23, Switzerland \\
and \\
Department of Physics and Astronomy, \\
University of Southern California, Los Angeles, CA 90089-0484}
\date{June 26, 1997}
\maketitle
\begin{abstract}
Type-IIB supersymmetric theories have an SL(2,Z) invariance, known as U-duality, 
which controls the non-perturbative behavior of the theory. Under SL(2,Z) the 
supercharges are doublets, implying that the bosonic charges would be singlets or
triplets. However, among the bosonic charges there are doublet strings and
doublet fivebranes which are in conflict with the doublet property of the supercharges. 
It is shown that the conflict is resolved by structure constants that depend
on moduli, such as the tau parameter, which transform under the same 
SL(2,Z). The resulting superalgebra encodes the non-perturbative duality properties of 
the theory and is valid for any value of the string coupling constant. 
The usefulness of the formalism is illustrated by applying it to purely algebraic 
computations of the tension of (p,q) strings, 
and the mass and entropy of extremal blackholes constructed from D-1-branes and
D-5-branes. In the latter case the non-perturbative coupling dependence of the BPS 
mass and renormalization is computed for the first time in this paper.
It is further argued that the moduli dependence of the superalgebra provides hints for 
four more dimensions beyond ten, such that the superalgebra is embedded in a 
fundamental theory which would be covariant under SO(11,3). An outline is
given for a matrix theory in 14 dimensions that would be consistent with 
M(atrix) theory as well as with the above observations.
\end{abstract}
\vskip2pc
]

\section{Extended type-IIB\ Supersymmetry and Duality}

Type-IIB\ supergravity and superstrings have two supercharges $Q_{\alpha
a}=(Q_{\alpha \left( 1\right) },Q_{\alpha \left( 2\right) }).$ The
16-component {\it real} spinors, $\alpha =1,\cdots ,16,$ have the same
ten-dimensional chirality. They are both the ${\bf 16}$ of SO$\left(
9,1\right) $. This is to be contrasted with type-IIA supercharges that are
one ${\bf 16}$ and one $\overline{{\bf 16}}$. In the {\it perturbative}
formulation of superstrings, the transformation rules of the dynamical
variables are such that, the 32 real supercharges obey the unextended
superalgebra 
\begin{equation}
\left\{ Q_{\alpha a},Q_{\beta b}\right\} =\delta _{ab}\,\gamma _{\alpha
\beta }^\mu \,P_\mu ,  \label{psusy}
\end{equation}
where $P_\mu $ is the momentum operator in ten dimensions, with $\mu
=0,1,\cdots ,9.$

The discovery of duality symmetries led to a larger picture in which strings
are in interaction with D-p-branes \cite{polchinski}\footnote{
An action of an open string interacting with zero branes (particles) at its
end points was suggested and studied a long time ago with a different
physical emphasis \cite{bh}. A re-examination of the approach in the context
of modern ideas may be enlightening.}. In this case the superalgebra must
have central extensions. The dynamical details of the underlying theory are
still being developed. However, independent of dynamical details, there is a
lot of information encoded in the superalgebra and we would like to take
maximum advantage of this information. The fully generalized superalgebra of
type-IIB in ten dimensions is 
\begin{eqnarray}
\left\{ Q_{\alpha a},Q_{\beta b}\right\} &=&\gamma _{\alpha \beta }^\mu
\,\,\left( p_\mu \delta _{ab}+e_\mu \tau _{ab}^3+d_\mu \tau _{ab}^1\right) 
\nonumber \\
&&+\gamma _{\alpha \beta }^{\mu _1\mu _2\cdots \mu _5}\,\left( 
\begin{array}{c}
x_{\mu _1\mu _2\cdots \mu _5}\delta _{ab} \\ 
+y_{\mu _1\mu _2\cdots \mu _5}\,\tau _{ab}^3 \\ 
+z_{\mu _1\mu _2\cdots \mu _5}\,\tau _{ab}^1
\end{array}
\right)  \label{gen} \\
&&+\gamma _{\alpha \beta }^{\mu _1\mu _2\mu _3}\,t_{\mu _1\mu _2\mu
_3}\,\left( i\tau ^2\right) _{ab}  \nonumber
\end{eqnarray}
where $\tau ^{1,2,3}$ are 2$\times 2$ Pauli matrices. This superalgebra is
meant to apply to any form of the theory, including the string
approximation, or the low energy approximation in the form of IIB
supergravity with all of its solitonic D-p-brane configurations. The
extensions $(e_\mu ,d_\mu )$ correspond to elementary strings and
D-1-strings respectively, $t_{\mu _1\mu _2\mu _3}$ corresponds to 3-branes,
and the self-dual tensors $x_{\mu _1\mu _2\cdots \mu _5},$ $y_{\mu _1\mu
_2\cdots \mu _5},$ $\,z_{\mu _1\mu _2\cdots \mu _5}$ correspond to three
types of self-dual 5-branes.

Altogether there are 528 bosonic operators, including the ``momentum'' $
p_\mu $. Other odd p-branes (p=-1,7,9) that can appear in type IIB theory
are related to the same set of 528 bosons after dualizing them. Note that we
have deliberately used the capital $P_\mu $ in (\ref{psusy}) and the low
case $p_\mu $ in (\ref{gen}). This is because the general relation between
the two involve moduli parameters in the non-perturbative theory, as will be
explained in the next section. This is why we used in quotes ``momentum'' in
referring to $p_\mu $. For the same reasons we should refer to $(e_\mu
,d_\mu )$ as ``strings'' in quotes. The true momentum $P_\mu $ and strings $
(E_\mu ,D_\mu )$ are linear combinations of $(p_\mu ,e_\mu ,d_\mu )$.

We will find that, in the general theory of type-IIB, the relation between
the two sets involves 6 functions of moduli that are Lorentz invariant in
10D. One of the moduli is the string coupling constant $g_s$ related to the
dilaton in the usual way. This will follow quite generally from the SL$
\left( 2,Z\right) _U$ U-duality properties of the type-IIB theory. It will
be demonstrated that in certain computations involving BPS sectors, only the
usual complex modulus $\tau $ enters in all expressions. Along the way we
will discover that the $\tau $ modulus together with at least one more real
modulus is needed to achieve SL$(2,Z)$ covariance of the superalgebra. We
will show that this symmetry and the moduli can be interpreted as if there
are 4 hidden dimensions with signature $\left( 2,2\right) ,$ and that three
(six) moduli can be constructed from one (two) SO$\left( 2,2\right) $ on
shell vector(s).

It is of interest in the present paper to determine the general dependence
of the 528 type-IIB bosons on the string coupling $g_s$ and other moduli in
any general theory that includes all D-p-branes. Recall first the form of
the superalgebra for elementary strings. In perturbative string theory,
without the D-p-branes, $g_s\rightarrow 0$, the only extensions in the
superalgebra are the winding numbers of the string $e_\mu \rightarrow E_\mu .
$ So, for elementary strings, at $g_s=0,$ the algebra has the form\footnote{
For an elementary string $P_\mu \pm E_\mu $ are the zero modes of the
left/right movers. Usually $E_\mu $ is taken zero in uncompactified
directions because of boundary conditions on closed or open strings. Its
non-vanishing components along compactified dimensions are determined by the
geometry of compactification. For example, if the 9$^{th}$ dimension is
compactified into a circle of radius $R_9$ then $E_9=nR_9,$ where $n$ is the
quantized winding number. In the general interacting case the boundary
conditions can change for open strings and $E_\mu $ can become non-zero in
all directions \cite{ibtokyo}. Since we are aiming at covering all
possibilities, we will formally allow all components of $E_\mu $. In can
then be specialized to the physical situation under discussion. The same
general remarks apply for all other elementary p-branes.} 
\begin{equation}
\lim_{g_s\rightarrow 0}\left\{ Q_{\alpha a},Q_{\beta b}\right\} =\gamma
_{\alpha \beta }^\mu \,\,\left( P_\mu \delta _{ab}+E_\mu \tau _{ab}^3\right)
.  \label{bry}
\end{equation}
If an elementary five-brane or elementary three brane is included in the
free theory, then one may add the corresponding extension that appears in (
\ref{gen}) at $g_s=0$. In this paper the free limit will be taken as a
boundary condition at $g_s=0$. As we have learned from past experience, if
one constructs an interacting action, $g_s\neq 0,$ based on the superalgebra
in eq.(\ref{psusy}) (e.g. type-IIB supergravity or superstrings), then one
finds out later, that for non-zero coupling and other non-zero moduli, the
theory has non-perturbative solitons in the form of D-p-branes. When all
p-branes are present the non-perturbative theory develops all the extensions
in the superalgebra (\ref{gen}), but with coefficients that depend on $g_s$
and the other moduli. It is our purpose here to find the general form of the
dependence on the moduli by taking advantage of the SL$\left( 2,Z\right) _U$
U-duality properties of the type-IIB theory. Of course, the general
superalgebra should be compatible with (\ref{gen}) but our interest is in
determining the non-perturbative moduli dependence of the various terms. The
form of the superalgebra that we will find at $g_s\neq 0$ through this
reasoning is expected to be valid in any formulation of the theory that
includes the D-p-branes in interaction with strings, and thus encodes a
certain amount of non-perturbative information independent of detailed
dynamics.

It will be important to note the maximal symmetries of the general type-IIB
superalgebra (\ref{gen}). It was first pointed out in \cite{stheory} that in
addition to 10D Lorentz symmetry SO$\left( 9,1\right) $, there is an SL$
\left( 2,R\right) _B$ symmetry in the type-IIB superalgebra (\ref{gen}),
including all extensions. Since the two supercharges are real they naturally
form a doublet $Q_{\alpha a}=(Q_{\alpha 1},Q_{\alpha 2})$ under an SL$\left(
2,R\right) _B$ transformation on the index $a=1,2.$ In the symmetric product
of the doublet there is a triplet and in the antisymmetric product there is
a singlet of SL$\left( 2,R\right) _B.$ Hence the superalgebra is written
covariantly as 
\begin{eqnarray}
\left\{ Q_{\alpha a},Q_{\beta b}\right\}  &=&\left( i\tau _2\tau _i\right)
_{ab}\left( \gamma _{\alpha \beta }^\mu \,\,p_\mu ^i+\gamma _{\alpha \beta
}^{\mu _1\mu _2\cdots \mu _5}\,x_{\mu _1\mu _2\cdots \mu _5}^i\right)  
\nonumber \\
&&\,+\left( i\tau _2\right) _{ab}\,\gamma _{\alpha \beta }^{\mu _1\mu _2\mu
_3}\,t_{\mu _1\mu _2\mu _3},  \label{nsusy}
\end{eqnarray}
where the triplet is associated with the adjoint representation of SL$\left(
2,R\right) _B,$ with Pauli matrices $\tau _i\equiv \left( -i\tau _2,\tau
_1,-\tau _3\right) .$ These are related to the symmetric matrices $i\tau
_2\tau _i=\left( 1,\tau _3,\tau _1\right) .$ Hence the ``momentum'' and
``strings'' are members of the SL$\left( 2,R\right) _B$ triplet 
\begin{equation}
p_\mu ^i=\left( p_\mu ,e_\mu ,d_\mu \right) .  \label{vec}
\end{equation}
The self-dual 5-branes also form a similar triplet, while the three brane is
a singlet.

The specialized form in (\ref{psusy}) is invariant under the subgroup SO$
\left( 2\right) \subset $SL$\left( 2,R\right) _B.$ The one in (\ref{bry}) is
invariant under the even smaller discrete subgroup $Z_2\subset $SO$\left(
2\right) $ $\subset $SL$\left( 2,R\right) _B$ if any component of $E_\mu $
is non-zero (and transforms as $E_\mu \rightarrow -E_\mu $ under the $Z_2).$
On the other hand if the D-1-brane charge is turned on, the full triplet $
p_\mu ^i$ is available to provide a basis for the full SL$\left( 2,R\right)
_B$ as in (\ref{nsusy}).

This symmetry has not received much attention so far. Is SL$\left(
2,R\right) _B$ related in any way to the SL$\left( 2,Z\right) _U$ of
U-duality \cite{ht}? At first sight there seems to be no relation, because
under U-duality the momentum $P_\mu $ is a singlet, while for SL$\left(
2,R\right) _B$ the ``momentum'' $p_\mu $ is a member of a triplet.
Similarly, the elementary and D-1-strings are SL$\left( 2,Z\right) _U$
doublets, while the ``strings'' that appear above are members of a SL$\left(
2,R\right) _B$ triplet. This poses a puzzle as to what is the significance
of the explicit SL$\left( 2,R\right) _B$ that stares at us in the
superalgebra? We will see later that by introducing moduli that transform
under both SL$\left( 2,Z\right) _U$ and SL$\left( 2,R\right) _B$ the
connection between the two will be made.

From another point of view SL$\left( 2,R\right) _B$ has been connected to
higher dimensions. In S-theory \cite{stheory} 3 extra dimensions beyond 10
were associated with SL$\left( 2,R\right) _B=\,$SO$\left( 2,1\right) ,$
(spacelike 11th \&13th and timelike 12th), but more recently in \cite{d14}
it was understood that SL$\left( 2,R\right) _B$ is more properly associated
with one of the SL$(2,R)$'s embedded in SO$\left( 2,2\right) =$SL$\left(
2,R\right) _B\otimes $ SL$\left( 2,R\right) _{B^{\prime }}$ where $\left(
2,2\right) $ corresponds to 2 timelike and two spacelike extra dimensions
beyond 10 (spacelike 11th \&13th and timelike 12th \& 14th). Indeed we will
see that the SL$(2,Z)_U$ that transforms the moduli is precisely the
discrete part of the SL$(2,R)_B$ embedded in SO$\left( 2,2\right) ,$ and
that a full set of 6 Lorentz invariant moduli, including the coupling
constant, are related to two ``on-shell'' vectors of SO$\left( 2,2\right) .$

In most of the paper we will concentrate on the type-IIB sector, and study
some non-perturbative properties of the general theory by taking advantage
of the SL$\left( 2,R\right) _B\longleftrightarrow $SL$\left( 2,Z\right) _U$
connection. We will construct the non-perturbative type-IIB\ superalgebra
with arbitrary values of the coupling constant and other moduli, show that
its non-perturbative duality properties under SL$\left( 2,Z\right) _U$ are
related to its hidden dimensions SO$\left( 2,2\right) $, and apply the
superalgebra to the derivation of non-perturbative results in order to
exhibit the utility and power of the approach. In particular, we will
compute in a few steps the tension of $(p,q)$ strings, and the mass spectrum
and entropy of black holes constructed from D-1-branes and D-5-branes.

The $\tau $ modulus that appears in such formulas, together with others
needed to achieve SL$\left( 2,Z\right) $ covariance, will be related to on
shell momentum-like vectors in 4 dimensions, beyond the usual ten
dimensions. This provides hints that in type-IIB supersymmetric theories
there is a hidden 14 dimensional structure. More generally, it was claimed
in \cite{d14} that a unified theory that is at least ten-dimensional Lorentz
covariant and is also covariant under type-IIA/IIB duality, is naturally a
14-dimensional theory with signature $\left( 11,3\right) ,$ with 64 real
supercharges. In this larger picture a constraint due to a gauge invariance
restricts the theory to live only in the short representation sectors of the
64 supercharges. This constraint is expressed as a 14D covariant BPS-like
master equation \cite{d14}. Dualities are naturally explained as symmetries
of the master equation. For example, it was discovered in S-theory that
type-IIA/IIB T-duality corresponds to flipping the 13th dimension with the
9th dimension \cite{stheory}\footnote{
This is easy to see this by ignoring the 14th dimension as in \cite{stheory}
. When compactified to 9 dimensions the type-IIA superalgebra with all 528
extensions has an SO$\left( 2,1\right) _A$ acting on the dimensions (9$^{th}$
,11$^{th}$,12$^{th}$)$_A$, while the type-IIB superalgebra has the SO$\left(
2,1\right) _B$ acting on the dimensions (13$^{th}$,11$^{th}$,12$^{th}$)$_B$.
The 13$^{th}-$9$^{th}$ flip is T-duality that maps these two into each other.
}. More will be said on this bigger picture in the last section of the
paper, where a general outline for a 14D matrix formulation that contains
dynamics, and which is consistent with the M(atrix) model of \cite{banks},
will be presented.

Finally, we remark that the methods of this paper can be extended to
U-duality covariant superalgebras in dimensions lower than ten, in
particular to $N=2$ in four dimensions, but this will not be pursued in the
present paper.

\section{Moduli, SL(2,Z) and 14D}

\subsection{Hints from supergravity}

The uniqueness of supergravity theories is established on the basis of local
supersymmetry. Accordingly, any type-IIB supersymmetric theory in 10
dimensions that contains the graviton, has type IIB supergravity as its low
energy limit in the massless sector. The massless spectrum of the
superstring contains the type-IIB\ supergravity supermultiplet consisting of
the graviton $g_{\mu \nu }$, two gravitinos $\psi _{\mu \alpha }^a$, two
2-index antisymmetric tensors $B_{\mu \nu }^a,$ one self--dual 4-index
antisymmetric tensor $A_{\mu \nu \lambda \sigma },$ two fermions $\chi
_\alpha ^a,$ and two scalars. The two scalars, identified as the dilaton $
\phi $ and the axion $a,$ are combined into the complex scalar field $\tau
=a+ie^{-\phi }.$ In the interacting supergravity theory, which corresponds
to the low energy limit of the interacting superstring theory, one notices
that there is an SL$(2,R)_U$ symmetry. Under SL$(2,R)_U$ the bosons
transform as follows: $g_{\mu \nu }$, $A_{\mu \nu \lambda \sigma }$ are
singlets, $B_{\mu \nu }^a$ form a doublet acting on the index $a=1,2$, and
the scalars undergo the fractional non-linear M\"{o}bius transformation 
\begin{equation}
\tau _g=\frac{a\tau +b}{c\tau +d},\quad g=\left( 
\begin{array}{ll}
a & b \\ 
c & d
\end{array}
\right) \in SL(2,R)_U.  \label{taut}
\end{equation}
This may be understood by taking the scalars to belong to the coset SL$
(2,R)_U/$SO$\left( 2\right) .$ The fermions $\left( \psi _{\mu \alpha
}^1,\psi _{\mu \alpha }^2\right) ,\left( \chi _\alpha ^1,\chi _\alpha
^2\right) $ transform linearly like doublets under an induced {\it gauged}
compact subgroup SO$\left( 2\right) \subset $SL$(2,R)$
\begin{equation}
\left( 
\begin{array}{ll}
cos\left( \Lambda \right) & sin\left( \Lambda \right) \\ 
-sin\left( \Lambda \right) & cos\left( \Lambda \right)
\end{array}
\right) \in SO\left( 2\right) .  \label{so2}
\end{equation}
The local gauge parameter of the transformation $\Lambda \left( x\right) $
depends on the global parameters $g\in $SL$(2,R)_U$ as well as on the local
scalar fields $\tau \left( x\right) .$ Thus, the fermions feel the full SL$
(2,R)_U$ transformation in the combination $\Lambda (g,\tau \left( x\right)
) $ that involves the scalar fields non-linearly.

The fermions are in one to one correspondence with the supercharges $\left(
Q_\alpha ^1,Q_\alpha ^2\right) $. Therefore the supercharges must also
transform as a doublet under the global part of SO$\left( 2\right) ,$ with a
constant parameter $\Lambda (g,\tau ),$ where $\tau ,$ called the modulus,
is the constant part of $\tau \left( x\right) $. This modulus is related to
the string coupling constant $g_s,$ and in lowest order perturbative regime
it takes the form 
\begin{equation}
\tau =\frac \theta {2\pi }+i\frac{4\pi ^2}{g_s^2}.  \label{tau}
\end{equation}
It transforms precisely as in (\ref{taut}). Thus, the supercharges of
supergravity feel the global SL$\left( 2,R\right) _U$ in the form of an
induced subgroup SO$\left( 2\right) $ that depends on the string coupling
constant. This is precisely the SO$\left( 2\right) $ symmetry of the
perturbative superalgebra (\ref{psusy}) discussed in the previous section.
Evidently this SO$\left( 2\right) _{U=B}$ subgroup is in the
non-perturbative SL$\left( 2,R\right) _B$ as seen in the previous section 
\begin{equation}
SO\left( 2\right) _U=SO\left( 2\right) _B\subset SL\left( 2,R\right) _B.
\end{equation}
This provides a hint for how to connect SL$\left( 2,R\right) _U$ and SL$
\left( 2,R\right) _B.$

In non-perturbative supergravity non-perturbative solitonic configurations
of the fields are permitted. To describe these field configurations one must
add more moduli (constants) corresponding to the boundary behavior of the
fields that are different in the perturbative versus the non-perturbative
theory. Effectively one introduces additional degrees of freedom through the
``back door'', as compared to the perturbative theory. The effect on the
superalgebra of the presence of the non-perturbative moduli is to turn on
central extensions, and hence change the supersymmetry properties of the
theory in the non-perturbative regimes. The theory is no longer the same,
and the supersymmetry transformation rules of the effective degrees of
freedom must be a ``new supersymmetry'' transformation rule, so that the
closure includes the central extensions. The extra charges together with
compactified components of the momentum form multiplets under the so called
U-duality transformations. Hence the momentum as well as the supercharges
cannot be immune to the SL$(2,R)_U$ transformations. This begins to smell
like the SL$\left( 2,R\right) _B\sim $ SL$\left( 2,R\right) _U$ symmetry.
But there seems to be conflicts in the transformation properties under SL$
\left( 2,R\right) _B$ versus SL$\left( 2,R\right) _U$ as described in the
previous section. The resolution of these conflicts involve the moduli and
their transformation properties as described below.

\subsection{Moduli dependent superalgebra}

The discussion above indicates that the nonperturbative supercharges and
their superalgebra depend generally on the moduli. This must be true since
they transform with the SO$\left( 2\right) _{U=B}$ whose parameters depend
on moduli. One of the moduli is the $\tau $ modulus (\ref{tau}) and we will
see that generally there are more. To emphasize this property we will write $
Q_{\alpha a}\left( \kappa \right) $ where $\left\{ \kappa \right\} $ is a
set of moduli. Since there is already dependence on moduli, is convenient to
choose a new basis for the $Q_{\alpha a}\left( \kappa \right) $ such that
the basis is related to the old one with moduli dependent coefficients. The
coefficients can be taken as functions of the coset SL$\left( 2,R\right) /$SO
$\left( 2\right) $ such that the $Q_{\alpha a}\left( \kappa \right) $
transform linearly under SL$\left( 2,R\right) =$SL$\left( 2,R\right) _{U=B}.$
That is, the modulus dependent SO$\left( 2\right) $ transformation is
compensated by the coefficients so that $Q_{\alpha a}\left( \kappa \right) $
transforms with modulus independent $g\in $SL$\left( 2,R\right) .$ One can
go back and forth between the two bases, so it will be convenient to use the
SL$\left( 2,R\right) $ covariant basis rather than the one that transforms
under the non-linear SO$\left( 2\right) $. Similarly, the operators that
appear in the general superalgebra (\ref{nsusy}) must be allowed to depend
on the moduli, and be consistent with the new basis. Hence in (\ref{nsusy})
one has $p_\mu ^i\left( \kappa \right) ,$ $x_{\mu _1\cdots \mu _5}^i\left(
\kappa \right) ,$ $y_{\mu _1\mu _2\mu _3}\left( \kappa \right) $ that are in
triplets and singlets of the linear SL$\left( 2,R\right) .$

We will now determine the moduli $\kappa $ on the basis of their SL$\left(
2,R\right) $ properties. They are Lorentz invariant in 10 dimensions, but
they transform non-trivially $\kappa \rightarrow \kappa _g$ under $g\in $SL$
\left( 2,R\right) $. In particular $\tau \rightarrow \tau _g$ must be given
by (\ref{taut}). In the new basis SL$\left( 2,R\right) _U$ transformations
are equivalent to applying SL$\left( 2,R\right) _B$ transformations (instead
of SO$\left( 2\right) )$ on the doublet indices $a$ on $Q_{\alpha a}\left(
\kappa \right) \,\,$or triplet indices $i$ on $p_\mu ^i\left( \kappa \right) 
$, $x_{\mu _1\cdots \mu _5}^i\left( \kappa \right) $. That is, 
\begin{eqnarray}
Q_{\alpha a}\left( \kappa \right) &\rightarrow &Q_{\alpha a}(\kappa
_g)=g_a^{\,\,\,b}Q_{\alpha b}(\kappa ),  \label{charges} \\
p_\mu ^i\left( \kappa \right) &\rightarrow &p_\mu ^i(\kappa
_g)=T_{\,\,\,\,j}^{\,i}(g)\,\,\,p_\mu ^j(\kappa ),\,\,etc.  \nonumber
\end{eqnarray}
where $T_{\,\,\,\,j}^{\,i}(g)$ is the triplet representation of $g\in
SL\left( 2,R\right) .$ Under these transformations we must seek a moduli
independent momentum operator $P_\mu $ that is a singlet of SL$\left(
2,R\right) _{U=B}$. Furthermore, one knows that the SL$\left( 2,R\right) _U$
doublet of supergravity fields $B_{\mu \nu }^a$ must couple to a doublet of
strings, the so called $(p,q)$ strings \cite{schwarz}. So we must also seek
a doublet of vectors $W_\mu ^a=(E_\mu ,D_\mu )$ representing the elementary
string (NS-NS sector) and the D-1-string (R-R sector). The three vectors $
p_\mu ^i\left( \kappa \right) $ that appear in the superalgebra must be a
combination of the moduli independent singlet and doublet vectors $P_\mu
,W_\mu ^a$ with coefficients that depend on the moduli $\kappa $. Hence we
must have 
\begin{equation}
p_\mu ^i\left( \kappa \right) =P_\mu \,\,v^i\left( \kappa \right)
+\varepsilon _{ab}W_\mu ^a\,\lambda ^{ib}\left( \kappa \right)
\end{equation}
where we have used the Levi-Civita symbol $\varepsilon _{ab}$ to combine
doublet indices into a singlet. We have introduced 3+3$\times 2=9$ functions 
$v^i\left( \kappa \right) ,\lambda ^{ib}\left( \kappa \right) $, but to
define the normalizations of $P_\mu ,W_\mu ^a$ we must normalize the
functions correctly. Thus we must find 9-3=6 constructed from moduli.
Consistency of the transformation properties (\ref{charges}) demand that
these functions satisfy 
\begin{equation}
v^i(\kappa _g)=T_{\,\,j}^i(g)\,v^j\left( \kappa \right) ,\quad \lambda
^{ia}(\kappa _g)=T_{\,\,j}^i(g)\,\,g_{\,\,\,b}^a\,\lambda ^{jb}\left( \kappa
\right) .  \label{tra}
\end{equation}
Then $P_\mu $ is a singlet and $W_\mu ^a$ is a doublet as desired 
\begin{equation}
P_\mu \,\rightarrow P_\mu ,\quad W_\mu ^a\,\rightarrow g_{\,\,\,b}^aW_\mu ^b.
\label{trans}
\end{equation}

It is convenient to rewrite the functions $v^i\left( \kappa \right) ,\lambda
^{ib}\left( \kappa \right) $ by converting the vector index $i$ into a pair
of symmetrized spinor indices, therefore we define the 9 functions $
G_{ab},\theta _{ab}^c$ 
\begin{equation}
G_{ab}\left( \kappa \right) \equiv \left( i\tau _2\tau _i\right)
_{ab}\,\,v^i\left( \kappa \right) ,\quad \theta _{ab}^c\left( \kappa \right)
\equiv \left( i\tau _2\tau _i\right) _{ab}\,\,\lambda ^{ic}\left( \kappa
\right) ,
\end{equation}
where the $ab$ indices are symmetric. The transformation laws (\ref{tra})
are equivalent to 
\begin{eqnarray}
G_{ab}\left( \kappa _g\right) &=&\left( gG\left( \kappa \right) g^T\right)
_{ab},\quad  \label{transf} \\
\theta _{ab}^c\left( \kappa _g\right) &=&g_{\,\,\,\,d}^c\left( g\theta
^d\left( \kappa \right) g^T\right) _{ab}.  \nonumber
\end{eqnarray}
In terms of these the general superalgebra (\ref{gen},\ref{nsusy}) takes the
form 
\begin{eqnarray}
\left\{ Q_{\alpha a},Q_{\beta b}\right\} &=&\gamma _{\alpha \beta }^\mu
\,\,\left( P_\mu G_{ab}+\varepsilon _{ab}W_\mu ^a\,\theta _{ab}^c\right) 
\nonumber \\
&&+\gamma _{\alpha \beta }^{\mu _1\mu _2\cdots \mu _5}\,\left( 
\begin{array}{c}
Y_{\mu _1\cdots \mu _5}\tilde{G}_{ab} \\ 
+\varepsilon _{dc}X_{\mu _1\cdots \mu _5}^d\,\tilde{\theta}_{ab}^c
\end{array}
\right)  \label{npsusy} \\
&&+\gamma _{\alpha \beta }^{\mu _1\mu _2\mu _3}\,T_{\mu _1\mu _2\mu
_3}\,\varepsilon _{ab}N\left( \kappa \right)  \nonumber
\end{eqnarray}
where, like the 1-branes, we have rewritten the three self-dual five branes $
x_{\mu _1\cdots \mu _5}^i\,$in terms of a singlet $Y_{\mu _1\cdots \mu _5}$
and a doublet $X_{\mu _1\cdots \mu _5}^d$ by introducing 9 functions $\tilde{
G},\tilde{\theta}$ that have identical transformation properties to those of 
$G,\theta .$ We have also introduced an overall normalization function $
N\left( \kappa \right) $ in front of the 3-brane $T_{\mu _1\mu _2\mu _3}$.
This function is a scalar under SL$\left( 2,R\right) $ transformations $
N\left( \kappa \right) \rightarrow N\left( \kappa _g\right) .$ We have taken
the pairs $G,\theta $ and $\tilde{G},\tilde{\theta}$ to be different in
general, but they may be the same depending on the details of the theory.
Similarly $N\left( \kappa \right) $ may be just a constant.

In this way we have shifted all the moduli dependence on the coefficients $
G,\theta ,\tilde{G},\tilde{\theta},N$ and identified the various operators
that are independent of the moduli. When SL$\left( 2,R\right) _U$
transformations are applied $\kappa \rightarrow \kappa _g$ it induces $
\kappa $-independent SL$\left( 2,R\right) _{B=U}$ transformations on the
operators, such that $\left( P_\mu ,Y_{\mu _1\cdots \mu _5},T_{\mu _1\mu
_2\mu _3}\right) $ behave as singlets and ($W_\mu ^a,X_{\mu _1\cdots \mu
_5}^a\,)$ behave as doublets$.$

\subsection{The functions}

Now, what are the functions that have these properties? We will construct $
G,\theta ,$ and assume that in the simplest case $\tilde{G},\tilde{\theta}$
are the same except possibly for overall factors similar to $N.$ Introduce
two 2$\times 2$ real matrices $A_{aa^{\prime }}\left( \kappa \right) $ and $
B_{aa^{\prime }}\left( \kappa \right) $ with determinants $
detA=1,\,\,detB=1, $ and postulate that the moduli $\kappa $ transform such
that these matrices transform only on the left side when $\kappa $ is
replaced by $\kappa _g$
\begin{equation}
A\left( \kappa _g\right) =gA\left( \kappa \right) ,\quad B\left( \kappa
_g\right) =gB\left( \kappa \right) ,  \label{left}
\end{equation}
Recall that after taking into account normalizations, taken together $
G_{ab}\left( \kappa \right) $ and $\theta _{ab}^c\left( \kappa \right) $
contain 6 independent functions of the moduli. On the other hand $A,B$ taken
together also have 6 independent functions of the moduli, so we may expect
to construct the most general $G_{ab}\left( \kappa \right) $ and $\theta
_{ab}^c\left( \kappa \right) $ in terms of $A\left( \kappa \right) $ and $
B\left( \kappa \right) .$ We can now give the 9 functions that have the
desired transformation properties 
\begin{eqnarray}
G_{ab}\left( \kappa \right) &=&\left( A1A^T\right) _{ab}\,\,\,,\quad
\label{gt} \\
\theta _{ab}^c\left( \kappa \right) &=&\left( A\tau ^1A^T\right)
_{ab}B_{c1}+\left( A\tau ^3A^T\right) _{ab}B_{c2}\,\,.  \nonumber
\end{eqnarray}
Next we construct $A,B$ for the general case. Each 2$\times 2$ real matrix $
A,B$ that transforms only on the left as in (\ref{left}) is really
equivalent to the vector representation $\left( 1/2,1/2\right) $ of SL$
\left( 2,R\right) _B\otimes $SL$\left( 2,R\right) _{B^{\prime }}=$SO$\left(
2,2\right) ,$ where SL$\left( 2,R\right) _{B^{\prime }}$ is not activated by
the SL$\left( 2,R\right) _{B=U}$ transformation of the moduli $\kappa
\rightarrow \kappa _g$ as seen in (\ref{left}). Therefore $A,B$ can be
parametrized by two ``on shell'' vectors $u^m,p^m$ of SO$\left( 2,2\right) $
by expanding each matrix in a basis of Pauli matrices$\,\,\,\tau _m=\left(
\tau _1,i\tau _2,\tau _3,1\right) ,$ i.e. 
\begin{equation}
B_{ab^{\prime }}=p^m\left( \tau _m\right) _{ab^{\prime }}\,\,,\quad \det
B=p^m\,\,p^n\eta _{mn}=1,
\end{equation}
where the signature $\eta _{mn}=\left( +,-,+,-\right) .$ Similar expressions
hold for $A\left( u\right) .$ The moduli $u^m$ and $p^m$ are generally
independent, but they could be related to each other in a special corners of
the theory. In particular, in the supergravity limit they are probably
related since there is only a single $\tau $ modulus in that case. In this
paper we will show that for computations in BPS sectors $A\left( u\right) $
disappears, so we will not need to know any of the possible relations
between the $u^m$ and $p^m$ moduli. Furthermore, for BPS sectors, only one
combination of the $p^m$ corresponding to the $\tau $ will survive, but the
remaining components must appear in more general computations. We will next
see how the $\tau $ modulus is embedded in SO$\left( 2,2\right) $.

Following the ideas of S-theory \cite{d14} it is tempting to interpret the
index $m$ as corresponding to two timelike and two spacelike dimensions
beyond the 10D. Hence, we will label $m=11,12,13,14.$ The explicit matrices
are expressed in terms of the lightcone type combinations of the components
of $p^m$ 
\begin{equation}
B=\left( 
\begin{array}{ll}
p_2^{14}+p_2^{13} & p_2^{11}+p_2^{12} \\ 
p_2^{11}-p_2^{12} & p_2^{14}-p_2^{13}
\end{array}
\right) .
\end{equation}
According to SL$\left( 2,R\right) _U$ transformations of eq.(\ref{left})
each column of this matrix transforms as a doublet. The same holds for the
two columns of $A\left( v\right) .$ We may now define the $\tau $ modulus in
terms of these. Define a complex doublet by combining the two columns of $B$ 
\begin{equation}
\left( 
\begin{array}{l}
z_1 \\ 
z_2
\end{array}
\right) =\left( 
\begin{array}{l}
\left( p_2^{14}+p_2^{13}\right) -i\left( p_2^{11}+p_2^{12}\right) \\ 
\left( p_2^{11}-p_2^{12}\right) -i\left( p_2^{14}-p_2^{13}\right)
\end{array}
\right)
\end{equation}
The ratio $\tau \equiv z_1/z_2$ undergoes the M\"{o}bius transformation
given in eq.(\ref{taut}). Note that the determinant (or ``on shell'')
condition for $B$ is equivalent to $
\mathop{\rm Im}
\left( z_1z_2^{*}\right) =1,$ or to $
\mathop{\rm Im}
\left( \tau \right) =\left| z_2\right| ^{-2},$ which requires $\tau $ to be
in the upper half plane. Thus, the complex modulus $\tau $ plus the phase of 
$z_2$ are 3 moduli equivalent to the matrix $B.$ We can write 
\begin{equation}
\left( 
\begin{array}{l}
z_1 \\ 
z_2
\end{array}
\right) =\frac{e^{i\phi }}{\sqrt{
\mathop{\rm Im}
\tau }}\left( 
\begin{array}{l}
\tau \\ 
1
\end{array}
\right)
\end{equation}
and 
\begin{equation}
B\left( \tau ,\phi \right) =\left( 
\begin{array}{ll}
\sqrt{
\mathop{\rm Im}
\tau } & \frac{
\mathop{\rm Re}
\tau }{\sqrt{
\mathop{\rm Im}
\tau }} \\ 
0 & \frac 1{\sqrt{
\mathop{\rm Im}
\tau }}
\end{array}
\right) \left( 
\begin{array}{ll}
-\sin \phi & -\cos \phi \\ 
\cos \phi & -\sin \phi
\end{array}
\right)
\end{equation}
Thus, the moduli correspond to an SO$\left( 2,2\right) $ vector $p_2^m\,$
that takes various values as a function of the coupling constant $g_s$ and
other moduli, while maintaining the SO$\left( 2,2\right) \,$ ``on shell''
constraint $p_2\cdot p_2=1.$

A similar statement can be made about $A.$ From the structure of (\ref{gt})
it is evident that $A$ corresponds to an SL$\left( 2,R\right) $ freedom for
changing the basis labelled by $a$ in $Q_{\alpha a}\left( \kappa \right) .$
This is related to the freedom of choosing the basis, as discussed in the
beginning of section 2b. This freedom can be used to relate the $u^m$ and $
p^m$ moduli or keep them independent depending on the particular corner of
the theory that one may wish to explore. For example, one may relate the
heterotic sector to the IIB theory by adjusting the $u^m$ moduli relative to 
$p^m$ moduli and taking appropriate limits so that the heterotic
superalgebra emerges from the type-IIB superalgebra discussed above.

For general values of the moduli we have obtained the non-perturbative form
of the superalgebra (\ref{npsusy}). In addition we have gained the following
perspective: The SL$\left( 2,R\right) _{U=B}$ transformations are a subset
of Lorentz transformations in the $\left( 2,2\right) $ space of the extra
dimensions. Lorentz transformations are equivalent to making the $\kappa
\rightarrow \kappa _g$ transformations on the moduli, including 
\begin{equation}
e^{i\phi _g}=e^{i\phi }\frac{\left( c\tau +d\right) }{\left| \left( c\tau
+d\right) \right| }\,.
\end{equation}
in addition to (\ref{taut}), and similarly for the moduli that parametrize $
A $. Rewriting the moduli in terms of the components of the SO$\left(
2,2\right) $ vectors brings out the 14D structure of the theory 
\begin{eqnarray}
\tau &=&\frac{(p_2^{14}+p_2^{13})-i(p_2^{12}+p_2^{11})}{
(p_2^{12}-p_2^{11})-i(p_2^{14}-p_2^{13})} \\
\tan \phi &=&(p_2^{14}-p_2^{13})/(p_2^{12}-p_2^{11}).  \nonumber
\end{eqnarray}

Up to now we have referred to continuous transformations $g\in $ SL$\left(
2,R\right) ,$ such as those that occur in (\ref{taut},\ref{transf}). Recall
that $g$ also acts on the doublet $W_\mu ^a.$ These operators have discrete
eigenvalues (the winding numbers) in the compactified dimensions. Since the
SL$\left( 2\right) $ transformation acts on a space of quantized eigenvalues
only the discrete subgroup SL$\left( 2,Z\right) $ remains as a symmetry in
the {\it compactified} theory.

It is worth asking the question if our construction has any relation to
F-theory \cite{vafa}? We do not know the answer, but one should note that
instead of defining a torus (which involved a worrisome analytic
continuation in F-theory), we have introduced SO$\left( 2,2\right) $ and
noticed four hidden dimensions rather than two. It may be interesting to
review F-theory constructions to see if they can be reinterpreted in terms
of a 14 dimensional formalism.

\section{Nonperturbative computations}

Certain properties of non-perturbative BPS states can be computed
algebraically by using the property that a BPS state corresponds to a short
representation of the superalgebra. In a short representation a subset of
supergenerators must vanish. If the superalgebra has the form 
\begin{equation}
\left\{ Q_{\alpha a},Q_{\beta b}\right\} =S_{\alpha a,\beta b},
\end{equation}
then in a short representation $S_{\alpha a,\beta b}$ must have zero
eigenvalues. This implies that the determinant is zero 
\begin{equation}
\det \left( S_{\alpha a,\beta b}\right) =0.
\end{equation}
This is a constraint that involves the momentum, the various bosonic charges
and the moduli. It must hold at any coupling since it is a purely algebraic
property. Hence all solutions of the determinant equation correspond to
non-perturbative results in the theory.

The determinant is invariant under similarity transformations. If $\tilde{G},
\tilde{\theta}$ are assumed to be proportional to $G,\theta $, then the
moduli $A\left( u\right) $ can be removed by a similarity transformation
consisting of the 32$\times 32$ matrix $1_{16}\otimes A\left( u\right) $.
Hence the moduli $u^m$ do not contribute in the BPS sector, and the matrix $
A $ that appears in (\ref{gt}) can be effectively gauge fixed to $A=1$ in
this sector.

In addition, one can make further orthogonal transformations on $S_{\alpha
a,\beta b}$ of the form $1_{16}\otimes T_2$ which do not change the gauge
fixed value of $A=1,$ since $T_21T_2^T=1.$ These are given by $T_2=\exp
\left( i\tau ^2\alpha \right) $, and they mix $\tau ^1$ with $\tau ^3.$ Then 
$\alpha $ can be chosen to remove the modulus $\phi $ from the expressions
in (\ref{gt}) provided we are in the BPS sector. There remains only the
complex modulus $\tau $ as the only gauge independent parameter in the BPS
sector.

These last two paragraphs explain why $\tau $ appears as the only
nonperturbative parameter in the following BPS calculations, although we
will include the most general form, including $\phi ,A\left( u\right) $
without choosing the gauges above. In non-BPS sectors one does not have the
gauge freedom described above, so at least the $\phi $ modulus must play a
role in the general sector. In addition the $u^m$ moduli are probably
important in general. Understanding their role may shed light on 14
dimensions.

\subsection{Tension of (q$_1$,q$_2$) Strings}

The ($q_1,q_2)$ strings are strings that carry $q_1$ units of NS-NS winding
numbers and $q_2$ units of R-R winding numbers, for $q_1,q_2$ relatively
prime integers. They occur as BPS states when the 9th dimension is
compactified on a circle of radius $R_9$. To compute the string tension we
set to zero all 5-branes $Y_{\mu _1\cdots \mu _5},X_{\mu _1\cdots \mu
_5}^a\,=0\,\,\,$and 3-branes $T_{\mu _1\mu _2\mu _3}=0$, and keep only the
Kaluza-Klein momentum $P_9=n/R_9$, and the 9th components of the two strings 
$W_9^a=m^aR_9$ which represent the winding numbers $m^a=(m_1,m_2)$ of the
NS-NS and R-R strings (or elementary string and D-1-string). The momentum in
9-dimensions $P_\mu $ is also present, but using the 9D Lorentz symmetry,
the momentum can be taken at rest $P_\mu =\left( M,\vec{0}\right) .$ On such
states the superalgebra takes the form 
\begin{equation}
S_{\alpha a,\beta b}=\gamma _{\alpha \beta }^0\,MG_{ab}+\gamma _{\alpha
\beta }^9\left( P_9G_{ab}+W_9^1\theta _{ab}^2-W_9^2\theta _{ab}^1\right) 
\end{equation}
where we insert the general expressions of (\ref{gt}). Recall that $\gamma
^0=1_{16}$ in the Weyl sector. For convenience we may choose a diagonal $
\gamma ^9=\sigma _3\otimes 1_8$. In computing the determinant, $A\left(
u\right) $ drops out since it has the form of a similarity transformation.
Then one can replace $G\rightarrow 1,$ $\theta ^2\rightarrow B^{21}\tau
_1+B^{22}\tau _3,$ $\theta ^1\rightarrow B^{11}\tau _1+B^{12}\tau _3$ and
compute the determinant easily 
\begin{eqnarray}
\det  &=&\left[ \left( M+P_9\right) ^2-W_9^a\left( B^{-1T}B^{-1}\right)
_{ab}W_9^b\right] ^8 \\
&&\times \left[ \left( M-P_9\right) ^2-W_9^a\left( B^{-1T}B^{-1}\right)
_{ab}W_9^b\right] ^8.  \nonumber
\end{eqnarray}
For generic values of $P_9,W_9^a$ there are four distinct eigenvalues where
the determinant vanishes. The BPS mass corresponds to the largest
eigenvalue, because the anticommutator of each of the 32 supercharges must
be either positive or zero. Thus, 
\begin{equation}
M_{BPS}=\frac{\left| n\right| }{R_9}+R_9\sqrt{m^am^bg_{ab}\left( \tau
\right) },
\end{equation}
where
\begin{eqnarray}
g_{ab}\left( \tau \right)  &=&\left( B^{-1T}B^{-1}\right) _{ab},  \nonumber
\\
&=&\frac 1{
\mathop{\rm Im}
\tau }\left( 
\begin{array}{ll}
\,\,\,\,\,\,1 & -
\mathop{\rm Re}
\tau  \\ 
-
\mathop{\rm Re}
\tau  & \,\,\,\,\,\left| \tau \right| ^2
\end{array}
\right) .
\end{eqnarray}
For generic $n,m^a$, the multiplicity of the zero is 8, therefore 8
supercharges vanish, giving a BPS space of strings with 1/4 supersymmetry.
In general the integers $\left( m_1,m_2\right) $ are not relatively prime.
Factoring out the largest common factor 
\begin{equation}
\left( m_1,m_2\right) =m\left( q_1,q_2\right) ,
\end{equation}
one obtains
\begin{eqnarray}
M_{BPS} &=&\frac{\left| n\right| }{R_9}+R_9\left| m\right| T_{q_1,q_2}, 
\nonumber \\
T_{q_1,q_2}\left( \tau \right)  &=&\sqrt{q^aq^bg_{ab}\left( \tau \right) }.
\end{eqnarray}
where $q_1,q_2$ are relatively prime. $T_{q_1,q_2}$ is interpreted as the
tension of the $(q_1,q_2)$ string at any coupling $\tau $ in agreement with 
\cite{schwarz}. In particular the NS-NS string $q^a=(1,0)$ and R-R string $
q^a=(0,1)$ have tensions 
\begin{equation}
T_{1,0}=\frac 1{\sqrt{
\mathop{\rm Im}
\tau }},\quad T_{0,1}=\frac{|\tau |}{\sqrt{
\mathop{\rm Im}
\tau }}.
\end{equation}
By comparing to (\ref{tau}) we see that in the weak coupling limit $\sqrt{
\mathop{\rm Im}
\tau }\rightarrow g_s^{-1}.$ So that the mass of the R-R string $q^a=(0,1)$
goes to infinity and decouples from the perturbative theory. By contrast, at
strong coupling, the NS-NS string $q^a=(1,0)$ becomes infinitely massive and
decouples. These two limits are interchanged by S duality, which is an
element of SL$\left( 2,Z\right) .$

Under SL$\left( 2,Z\right) $ the quanta $m^a$ transform as a doublet and $
\tau $ transforms under M\"{o}bius transformations as in (\ref{taut}), while
the mass formula or the tension remains invariant. This invariance was built
in the superalgebra, so it is no surprise that it is present in the mass
formula derived through an SL$\left( 2,Z\right) $ covariant procedure. By
obtaining this well known result from the nonperturbative superalgebra we
have illustrated that the algebra encodes useful information.

\subsection{Mass and Entropy of BPS Extremal Black holes}

Consider supergravity as the low energy limit of any type-IIB theory
(superstring, M-theory, etc.). It is known that in supergravity there are
black hole solutions. At space infinity, far away from the black hole, the
field configurations for these solutions are proportional to p-brane
charges. These charges that define the black hole state are the ones that
appear in the superalgebra as central extensions. It is known that the
extremal black holes are BPS states, therefore many of their properties,
such as masses, can be determined by algebraic means. In particular, it is
known that the area of the horizon of the extremal black hole, which is
proportional to the entropy, is given as a U-duality invariant expression
constructed from the central extensions of the superalgebra \cite{kallosh}.
We can therefore apply a purely algebraic approach to compute the mass and
entropy of supersymmetric extremal black holes. With our formalism we can
compute these quantities at any value of the coupling.

As an example we consider a popular black hole configuration in 5-dimensions
that is constructed from $Q_1$ D-1-branes and $Q_5$ D-5-branes. It has the
following three charges \cite{maldacena}: Kaluza-Klein momentum $P_9=n/R_9,$
string winding $W_9^{\left( 2\right) }=Q_1R_9,$ D-5-brane winding $
X_{56789}^{\left( 2\right) }=Q_5R_9V_4.$ Here $n,Q_1,Q_5$ are quantized
integers of any sign (brane/anti-brane etc.), $R_9$ is the radius of the
circle for the 9th compactified dimension, and $R_9V_4$ is the volume of the
D-5 brane. There is also a momentum in 5-dimensions which is taken at rest $
P_\mu =\left( M,\vec{0}\right) ,$ with $M$ positive. All other extensions in
the superalgebra are set to zero for this black hole state. The
non-perturbative superalgebra takes the form 
\begin{eqnarray}
S_{\alpha a,\beta b} &=&\left( \gamma _{\alpha \beta }^0\,M+\gamma _{\alpha
\beta }^9P_9\right) G_{ab}-\gamma _{\alpha \beta }^9W_9^{\left( 2\right)
}\theta _{ab}^{\left( 1\right) }  \label{central} \\
&&-\gamma _{\alpha \beta }^{56789}X_{56789}^{\left( 2\right) }\theta
_{ab}^{\left( 1\right) }.  \nonumber
\end{eqnarray}
As in the previous subsection $A$ drops out in the computation of the
determinant, so we may replace $G\rightarrow 1,$ $\theta ^1\rightarrow
B^{11}\tau _1+B^{12}\tau _3.$ By making similarity transformations which do
not change the determinant of $S$, one can choose a new basis for the $\tau $
matrices such that $\theta ^1$ is diagonalized in the new form 
\begin{eqnarray}
\theta ^1 &\rightarrow &\tau _3\sqrt{\left( B^{11}\right) ^2+\left(
B^{12}\right) ^2}  \nonumber \\
&=&\tau _3\frac{\left| \tau \right| }{\sqrt{
\mathop{\rm Im}
\tau }}.
\end{eqnarray}
Recall that in the Weyl basis $\gamma ^0$ $=1_{16},$ furthermore $\gamma ^9$
and $\gamma ^{56789}$ commute with each other. So one may use the following
16$\times 16$ diagonal matrices to represent them 
\begin{equation}
\gamma ^9=1_4\otimes \sigma _3\otimes 1_2,\quad \gamma ^{56789}=1_4\otimes
1_2\otimes \sigma _3.  \label{gammas}
\end{equation}
In this form the 32$\times 32$ matrix $S_{\alpha a,\beta b}$ is diagonal.
For generic $M,P_9,W_9^{\left( 2\right) },X_{56789}^{\left( 2\right) }$ it
has eight distinct eigenvalues, each one being fourfold degenerate,
indicating that this configuration has 4 zero supercharges if the mass $M$
is chosen equal to the BPS mass. The relative signs must be chosen so that
the BPS mass is the largest eigenvalue, so that each of the 32 supercharges
has a positive anticommutator. Therefore we obtain only positive
contributions from each term 
\begin{eqnarray}
M_{BPS} &=&\left| P_9\right| +\left( \left| W_9^{\left( 2\right) }\right|
+\left| X_{56789}^{\left( 2\right) }\right| \right) \frac{\left| \tau
\right| }{\sqrt{
\mathop{\rm Im}
\tau }}  \nonumber \\
&=&\frac{\left| n\right| }{R_9}+\left( \left| Q_1\right| +V_4\left|
Q_5\right| \right) \frac{R_9\left| \tau \right| }{\sqrt{
\mathop{\rm Im}
\tau }}.
\end{eqnarray}
Since the zero eigenvalue of $S_{\alpha a,\beta b}$ is fourfold, four
supercharges vanish. Therefore this BPS spectrum of black holes has 1/8
(=4/32) supersymmetry at generic values of $n,Q_1,Q_5$.

This non-perturbative result, valid for all values of the coupling constant
and moduli, is more general than others available in the literature. It is
in agreement with \cite{maldacena} for the special values $
\mathop{\rm Re}
\tau =0$ and $\sqrt{
\mathop{\rm Im}
\tau }\sim 1/g_s$ valid at weak coupling. As demonstrated, it follows only
from algebraic properties of the superalgebra, plus physical insight into
the meaning of the various parameters that characterize the quantum numbers
of the black hole. Note that the non-perturbative tension of the D-1 brane
is consistent with the previous subsection.

The entropy of an extremal black hole in five dimensions is expressed
generally by $S=A_5/4G_5=2\pi \sqrt{\left| I_5\right| },$ where $G_5$ is the
Newton constant in 5 dimensions and $A_5$ is the area of the black hole.
This expression is given by the $E_{6,6}$ cubic invariant $I_5$ constructed
from the central extensions \cite{kallosh}\cite{ver}\cite{hc} 
\begin{equation}
I_5=C\,Tr\left( Z\Omega Z\Omega Z\Omega \right) ,  \label{ent5}
\end{equation}
where $\Omega $ is the Sp$\left( 8\right) $ invariant metric and $Z$
contains the 27 of $E_{6,6}$ in the form of an 8$\times 8$ antisymmetric
matrix of central extensions which satisfies Tr$\left( Z\Omega \right) =0.$
The factor $C$ in front accounts for a renormalization of the central
extensions, and it will be explained below. In our case the form of the
central extension $Z,$ in 10 dimensional units, is given by the matrix in (
\ref{central}) excluding the mass term. An 8$\times 8$ antisymmetric $Z$ is
formed by using the gamma matrices in (\ref{gammas}), excluding the first
factor $1_4$ (spinor space for the 5 dimensions) and replacing the diagonal $
\sigma _3$'s by antisymmetric $\sigma _2$'s. 
\begin{eqnarray}
Z &=&\,\sigma _2\otimes 1_2\otimes G\,\,P_9-\sigma _2\otimes 1_2\otimes
\theta ^{\left( 1\right) }W_9^{\left( 2\right) } \\
&&-1_2\otimes \sigma _2\otimes \theta ^{\left( 1\right) }X_{56789}^{\left(
2\right) }.  \nonumber
\end{eqnarray}
The 8$\times 8$ antisymmetric $\Omega $ orthogonal to $Z$ is $\Omega
=1_2\otimes \sigma _2\otimes G^{-1}$. Then $I_5$ reduces to 
\begin{equation}
I_5=24C\,\,Tr\left( \theta ^{\left( 1\right) }G^{-1}\theta ^{\left( 1\right)
}G^{-1}\right) P_9W_9^{\left( 2\right) }X_{56789}^{\left( 2\right) }. 
\nonumber
\end{equation}
This expression is simplified by using the property of the trace to
eliminate the matrix $A$, thus substituting again $G\rightarrow 1_2$ and$
\,\,\,\theta ^1\rightarrow B^{11}\tau _1+B^{12}\tau _3.$ The result is 
\begin{equation}
I_5=24C\frac{2\left| \tau \right| ^2}{
\mathop{\rm Im}
\tau }R_9V_4nQ_1Q_5=nQ_1Q_5,
\end{equation}
which gives the well known moduli independent entropy \cite{maldacena},
provided 
\begin{equation}
C=\frac{
\mathop{\rm Im}
\tau }{48R_9V_4\left| \tau \right| ^2}.
\end{equation}
The factor $C$ is non-trivial because we have expressed the central
extensions in 10 dimensional units, as they appear in the 10D superalgebra,
which is renormalized relative to the one in \cite{kallosh}\cite{ver}\cite
{hc}. Renormalization factors for the metric and fermions in 5D introduce a
renormalization of the 5D supercharges, and hence of the central extensions.
We have found that $C$ has non-perturbative contributions. 

We have demonstrated the usefulness of the approach by deriving new
non-perturbative results for the mass $M_{BPS}$ and for the factor $C$
related to renormalization.

Many more new non-perturbative results can be derived for other physical
states by following a similar procedure for any p-brane configuration.
Furthermore, since the superalgebra must be valid for matrix elements of
scattering states, there must be various sum rules and relations among
scattering amplitudes that can be derived algebraically by sandwiching the
superalgebra between scattering states. The methods for obtaining such
results are similar to the current algebra techniques used in strong and
weak interactions in the sixties. Such results will be valid
non-perturbatively for any value of the coupling constant since the
superalgebra is non-perturbative. It would be interesting to derive such
relations in the future.

\section{S-theory based on matrices in 14D}

Some very complicated arguments that yield non-perturbative results about
the theory seem to rely on dynamics, but the results shown above, and many
others, really depend on the properties of the superalgebra and would be the
same in any model with the same extended supersymmetry. As advocated in the
S-theory approach \cite{stheory}, it is useful to recognize the general
properties of the symmetry and distinguish the results that follow from it
independently from the dynamics. Of course, dynamical realizations of the
superalgebra provide greater insight into the physics. Using the general
algebraic properties as a guide to construct the dynamics is bound to be
fruitful. We will follow this route to suggest an outline for a dynamical
model below.

We sketch the basic elements of a new dynamical matrix construction \cite
{matrix} that realizes the bigger picture compatible with the general
superalgebra. Although the emphasis in this paper is not on this theory, a
brief outline is included here in order to provide a perspective for a more
concrete realization of the algebraic properties of S-theory discussed in
this paper and elsewhere \cite{stheory}\cite{d14}.

Let us first recall the general algebraic picture that goes beyond the
structures in the previous sections. It was claimed in \cite{d14} that the
unified theory that exhibits explicitly at least ten-dimensional Lorentz
covariance and type-IIA/IIB duality, is naturally a 14-dimensional theory
with signature $\left( 11,3\right) ,$ because it must be formulated in terms
of 64 real supercharges. Supersymmetry in higher than 11 dimensions deviates
substantially from the usual form. In 14D the general superalgebra, with 64
real supercharges corresponding to the Weyl-Majorana spinor, is given by 
\begin{equation}
\left\{ Q_\alpha ,Q_\beta \right\} =\gamma _{\alpha \beta }^{\mu _1\mu _2\mu
_3}T_{\mu _1\mu _2\mu _3}+\gamma _{\alpha \beta }^{\mu _1\cdots \mu
_7}Z_{\mu _1\cdots \mu _7}^{+}  \label{d14susy}
\end{equation}
where $Z_{\mu _1\cdots \mu _7}^{+}$ is self dual. As discussed in \cite{d14}
, for the theory to have effectively only 32 real supercharges, a gauge
symmetry must be built in such that it provides the constraint that this
supersymmetry must live only in the short multiplet sectors. A fully SO$
\left( 11,3\right) $ covariant equation that insures this property is the
master equation (that should follow from a gauge invariance) 
\begin{equation}
\det \left( \gamma ^{\mu _1\mu _2\mu _3}T_{\mu _1\mu _2\mu _3}+\gamma ^{\mu
_1\cdots \mu _7}Z_{\mu _1\cdots \mu _7}^{+}\right) =0.  \label{detcond}
\end{equation}
As explained in \cite{d14}, this BPS-like equation has four\footnote{
In \cite{d14} the fourth branch was missed. I thank S. Yankielowicz for
pointing it out.} main branches of solutions falling into sectors $A,B,C,D$
characterized by the maximal isometries of their superalgebras (if certain
moduli are frozen, see \cite{d14}) 
\begin{eqnarray}
A &:&SO\left( 10,2\right) \otimes SO\left( 1,1\right) ,  \nonumber \\
B &:&SO\left( 9,1\right) \otimes SO\left( 2,2\right) ,  \label{branches} \\
C &:&SO\left( 8,0\right) \otimes SO\left( 3,3\right) ,  \nonumber \\
D &:&SO\left( 6,2\right) \otimes SO\left( 3,1\right) \otimes SO\left(
1,1\right) .  \nonumber
\end{eqnarray}
Each branch has 32 real supercharges, and contains sub-branches with fewer
supercharges. The branches labelled by $A,B,$ when viewed from the point of
view of 10 dimensions, have precisely the maximally extended superalgebras
of types-IIA,IIB in ten dimensions. The $C,D$-branches do not have 10D
covariance but have 4D covariance since they contain SO$\left( 3,1\right) $
as a subgroup$.$ Sub-branches with less supersymmetry, corresponding to
heterotic, type-I, and all compactifications, are all solutions of the same
14D master equation. The symmetries of the master equation (similarity
transformations on the 64$\times 64$ matrix in spinor space) permit
transformations from one solution to another, and these correspond to
dualities \cite{d14}.

Now we present some initial ideas for building a dynamical model compatible
with the general algebraic picture. It is based on an SO$\left( 11,3\right) $
covariant infinite dimensional matrix model that borrows from the current
ideas in matrix models for M-theory \cite{banks}. Suppose there is an action
for a matrix model in 14 dimensions, with 14 bosonic Hermitian matrices $
X_{ij}^\mu ,$ $\mu =1,\cdots ,14,$ with signature $\left( 11,3\right) ,$ and
64 fermionic Hermitian matrices $\theta _{ij}^\alpha ,$ $\alpha =1,\cdots
,64,$ in the Weyl-Majorana spinor representation of SO$\left( 11,3\right) .$
Since supersymmetry in higher than 11 dimensions deviates substantially from
the usual form, as seen in (\ref{d14susy}), the 14D matrix action cannot
closely resemble the reduced Super Yang Mills action in 10 dimensions.
However, the superalgebra constructed from the matrices is expected to have
extensions that are similar to the ones found in 10D matrix models. Namely,
consider the form 
\begin{eqnarray}
\left\{ Q_\alpha ,Q_\beta \right\} &=&\gamma _{\alpha \beta }^{\mu _1\mu
_2\mu _3}\left[ \frac 1NTr(X_{\mu _1}X_{\mu _2}X_{\mu _3})+\cdots \right] \\
&&+\gamma _{\alpha \beta }^{\mu _1\cdots \mu _7}\left[ \frac 1NTr(X_{\mu
_1}X_{\mu _2}\cdots X_{\mu _7})+\cdots \right] ,  \nonumber
\end{eqnarray}
where the extensions constructed from matrices 
\begin{equation}
T_{\mu _1\mu _2\mu _3}=\frac 1{2N}Tr(X_{\mu _1}X_{\mu _2}X_{\mu _3}-X_{\mu
_2}X_{\mu _1}X_{\mu _3})+\cdots ,
\end{equation}
and similarly for $Z_{\mu _1\cdots \mu _7}^{+}$, are completely
antisymmetric in the 14D Lorentz indices.

By now it is known that infinite dimensional matrices can be chosen to
correspond to any collection of p-branes, with several types of p-branes
appearing simultaneously. In our case we require that their configurations
obey the master equation described above.

It is evident that all known p-brane solutions in 10D matrix models \cite
{banks} form a subset of solutions to our equations since they can be
embedded in 14D as in \cite{d14}. For example one can take 10 matrices as in 
\cite{banks} and 4 other constant matrices that point along constant
orthogonal vectors embedded in SO$\left( 2,2\right) .$ The determinant
condition (\ref{detcond}) is a restriction on the constant vectors such that
the solution for $T_{\mu _1\mu _2\mu _3},\,Z_{\mu _1\cdots \mu _7}^{+}$
falls into one of the branches or sub-branches in (\ref{branches}).

This approach, which is compatible with other current ideas on matrix
models, contains dynamics in a duality and Lorentz covariant formalism,
while providing a concrete realization of the algebraic properties of
S-theory described in this paper and elsewhere \cite{stheory}\cite{d14}. The
bigger picture outlined here schematically will be explored elsewhere in
greater detail \cite{matrix}.

\section{Acknowledgments}

I would like to thank Costas Kounnas for very valuable conversations that
provided hints for connecting the SO(2,2) dimensions to moduli. I also
acknowledge useful conversations with S. Yankielowicz, J. Maldacena, S.
Ferrara, P. Townsend, A. Tseytlin, M. Cvetic, and T. Banks. The research was
partially supported by a DOE grant No. DE-FG03-84ER40168.

\medskip

\vfill\eject 

\end{document}